\documentclass[12pt]{article}
\usepackage{epsf,epsfig,graphics}
\newcommand{\titul}[1] {\begin{center}{\Large {\bf #1 } } \end{center}
\vskip 0.8cm}

\newcommand{\autor}[1] {\begin{center}  {\bf \lineskip .3cm #1  }
                      \end{center} }

\newcommand{\lugar}[1] {\begin{center}  {\normalsize \bf \it #1   }
\end{center}}
\newcounter{muni}

\begin{document}
\hbadness=10000 \pagenumbering{arabic}
\begin{titlepage}
%
%\prepr{hep-ph/0308xxx \\IPAS-03-05 }

\titul{\bf Weak phases from the $B\to \pi\pi$, $K\pi$ decays}

\autor{Yeo-Yie Charng$^1$\footnote{Email:
charng@phys.sinica.edu.tw} and Hsiang-nan
Li$^{1,2}$\footnote{Email: hnli@phys.sinica.edu.tw}}
\lugar{$^1$Institute of Physics, Academia Sinica, Taipei,\\
Taiwan 115, Republic of China}
\lugar{$^2$Department of Physics, National Cheng-Kung University,\\
Tainan, Taiwan 701, Republic of China}

\vskip 2.0cm {\bf PACS index : 13.25.Hw, 11.10.Hi, 12.38.Bx,
13.25.Ft}

\thispagestyle{empty}
%%%%%%%%%%%%%%%%%%%%%%%%%%%%%%%%%%%%%%%%%%%%%%%%%%%%%%%%%%%%%%%%%%%%%%%%%%%%%
%\newpage
\vspace{10mm}
\begin{abstract}
Recent data of two-body nonleptonic $B$ meson decays allow a
topological-amplitude analysis up to the $O(\lambda^2)$ accuracy,
where $\lambda$ denotes the Wolfenstein parameter. We find an
exact solution from the $B\to\pi\pi$ data and an exact solution
from the $B\to K\pi$ data, which satisfy the approximate SU(3)
flavor symmetry. These solutions indicate that the
color-suppressed tree amplitude is large, all other amplitudes can
be understood within the standard model, and the weak phases
$\phi_2\approx 90^o$ and $\phi_3\approx 60^o$ are consistent with
the global unitarity triangle fit.

%Our conclusions are (i)
%final-state interaction may not be important; (ii) a
%color-suppressed tree amplitude may be larger than expected; (iii)
%an electroweak penguin contribution may be as small as expected;
%(iv) a large strong phase between the tree and penguin amplitude
%is confirmed; (v) tree and electroweak penguin amplitudes have
%similar strong phases.
\end{abstract}
\thispagestyle{empty}
\end{titlepage}

To determine the weak phases in the Kobayashi-Maskawa ansatz for
CP violation \cite{KoMa}, one either resorts to theoretically
clean modes, which are usually experimentally difficult, or to the
modes with higher feasibility, which, however, require theoretical
inputs \cite{Charng}. Recently, we have adopted the
topological-amplitude parametrization for two-body nonleptonic $B$
meson decays, in which the theoretical inputs are the counting
rules for various decay amplitudes \cite{CC} in terms of powers of
the Wolfenstein parameter $\lambda\sim 0.22$. These counting rules
are supported by the known QCD theories \cite{KLS,BBNS,KL,CKL},
and slightly different from those postulated in \cite{GHL}. The
strategy of this method is to drop the topologies with higher
powers of $\lambda$ until the number of free parameters are equal
to the number of available measurements. The weak phases and the
amplitudes are then solved exactly by comparing the resultant
parametrization with experimental data. Afterwards, it should be
examined whether the obtained amplitudes obey the power counting
rules. If they do, the extracted weak phases suffer only the
theoretical uncertainty from the neglected topologies. If not, the
inconsistency could be regarded as a warning to the QCD theories.

Because the data were not complete, the analysis performed in
\cite{Charng} was limited to the $O(\lambda)$ accuracy: the
electroweak penguin amplitude $P_{ew}$ has been neglected for the
$B\to \pi\pi$ decays. The color-suppressed tree amplitude $C$, the
color-suppressed electroweak penguin amplitude $P_{ew}^c$, and the
tree annihilation amplitude $T^a$ have been neglected for the
$B\to K\pi$ decays. In this work we shall improve the accuracy up
to $O(\lambda^2)$, since recent experimental progress has allowed
this study. Moreover, we shall look for the solutions, in which
the amplitudes of each topology from the $B\to \pi\pi$, $K\pi$
modes are consistent with the approximate SU(3) flavor symmetry.
If such solutions exist (there is no guarantee for the existence
in this method), all the above data can be understood in a
consistent way, and the determination of the weak phases $\phi_2$
and $\phi_3$ will be convincing. The $B\to \pi\pi$, $K\pi$ old
data have been investigated in \cite{BFRPR,WZ} based on the SU(3)
flavor symmetry to some extent, and an extracted large $P_{ew}$
has been claimed to signal new physics. We shall point out that
the large $P_{ew}$ is a consequence of the {\it strong} assumption
of the SU(3) flavor symmetry and of the old data. Different
prescriptions for taking into account SU(3) symmetry breaking
effects have led to different extractions of amplitudes
\cite{BFRPR,WZ,CGRS}, while our exact solutions avoid this
ambiguity. As shown below, the new data in fact imply only a large
$C$, and all other amplitudes, including $P_{ew}$, can be
understood within the standard model.

The most general topological-amplitude parametrization of the
$B\to\pi\pi$ decay amplitudes is written as
\begin{eqnarray}
\sqrt{2}A(B^+\to \pi^+\pi^0)&=&-T\left[1+\frac{C}{T}
+\frac{P_{ew}}{T}e^{i\phi_2}\right]\;,
\nonumber\\
A(B_d^0\to \pi^+\pi^-)&=&-T\left(1
+\frac{P}{T}e^{i\phi_2}\right)\;,
\nonumber\\
\sqrt{2}A(B_d^0\to \pi^0\pi^0)&=&T\left[\left(
\frac{P}{T}-\frac{P_{ew}}{T}\right)
e^{i\phi_2}-\frac{C}{T}\right]\;, \label{Mbpi1}
\end{eqnarray}
with the power counting rules,
\begin{eqnarray}
& &\frac{P}{T}\sim \lambda\;,\;\;\;\; \frac{C}{T}\sim \lambda\;,
\;\;\;\; \frac{P_{ew}}{T}\sim \lambda^2\;. \label{po}
\end{eqnarray}
We have adopted the $t$-convention for the above parametrization
with the product of the Cabibbo-Kobayashi-Maskawa (CKM) matrix
elements $V_c=V_{cd}V_{cb}^*$ being eliminated by virtue of the
unitarity relation. In this convention the tree amplitudes contain
the weak phase $\phi_3$, and the penguin amplitudes contain
$\phi_1$. $\phi_3$ is then factored out, such that the penguin
amplitudes carry the weak phase $\phi_2=180^o-\phi_1-\phi_3$
eventually. There are 4 independent amplitudes, namely, 7
parameters, because an overall phase can always be removed. Including
the weak phase $\phi_2$, there are 8 unknowns in Eq.~(\ref{Mbpi1}). 
The available data
of the branching ratios and the CP asymmetries are summarized as
\cite{HFAG},
\begin{eqnarray}
& &{\rm Br}(B^\pm\to \pi^\pm\pi^0) =(5.5\pm 0.6)\times
10^{-6}\;\;({\rm updated})\;,
\nonumber\\
& &{\rm Br}(B_d^0\to \pi^\pm\pi^\mp) =(4.6\pm 0.4)\times
10^{-6}\;,
\nonumber\\
& &{\rm Br}(B_d^0\to \pi^0\pi^0) =(1.51\pm 0.28)\times
10^{-6}\;\;({\rm updated})\;,
\nonumber\\
& &{\cal A}(B_d^0\to \pi^\pm\pi^\mp)=(37\pm 11)\%\;\;
({\rm updated})\;,\nonumber\\
& &{\cal S}(B_d^0\to \pi^\pm\pi^\mp)=-(61\pm 14)\%\;\;({\rm
updated})\;,
\nonumber\\
& &{\cal A}(B^\pm\to \pi^\pm\pi^0) =-(1\pm 7)\%\;\;({\rm new})
\;, \nonumber\\
& &{\cal A}(B_d^0\to \pi^0\pi^0) =(28\pm 39)\%\;\;({\rm new})
\;.\label{pp}
\end{eqnarray}

Following \cite{Charng}, the parametrization for the $B\to K\pi$
decays is written, up to $O(\lambda^2)$, as
\begin{eqnarray}
A(B^+\to K^0\pi^+)&=&P'\;,\nonumber\\
A(B_d^0\to
K^+\pi^-)&=&-P'\left(1+\frac{T'}{P'}e^{i\phi_3}\right)\;,
\nonumber\\
\sqrt{2}A(B^+\to K^+\pi^0)&=&-P'\left[1+\frac{P'_{ew}}{P'}
+\left(\frac{T'}{P'}+\frac{C'}{P'} \right)e^{i\phi_3}\right]\;,
\nonumber\\
\sqrt{2}A(B_d^0\to K^0\pi^0)&=&P'\left(1 -\frac{P'_{ew}}{P'}
-\frac{C'}{P'}e^{i\phi_3}\right)\;, \label{Mbpp1}
\end{eqnarray}
with the power counting rules,
\begin{eqnarray}
\frac{T'}{P'}\sim  \lambda\;,\;\;\;\;
\frac{P'_{ew}}{P'}\sim\lambda\;,\;\;\;\;
\frac{C'}{P'}\sim\lambda^2\;. \label{pow}
\end{eqnarray}
There are also 8 unknowns including the weak phase $\phi_3$, which
will be solved from the 8 experimental inputs \cite{HFAG},
\begin{eqnarray}
& &{\rm Br}(B^\pm\to K^0\pi^\pm)=(24.1\pm 1.3)\times
10^{-6}\;\;({\rm updated})\;,
\nonumber\\
& &{\rm Br}(B_d^0\to K^\pm\pi^\mp) =(18.2\pm 0.8)\times 10^{-6}\;,
\nonumber\\
& &{\rm Br}(B^\pm\to K^\pm\pi^0)=(12.1\pm 0.8) \times
10^{-6}\;\;({\rm updated})\;,
\nonumber\\
& &{\rm Br}(B_d^0\to K^0\pi^0)=(11.5\pm 1.0) \times 10^{-6} \;,
\nonumber\\
& &{\cal A}(B_d^0\to K^\pm\pi^\pm)=-(11.3\pm 1.9)\%\;\;({\rm
updated})\;,
\nonumber\\
& &{\cal A}(B^\pm\to K^\pm\pi^0)=(4\pm 4)\%\;\;({\rm updated})\;,
\nonumber\\
& &{\cal A}(B_d^0\to K^0\pi^0)=(9\pm 14)\%\;\;({\rm new})\;,
\nonumber\\
& &{\cal S}(B_d^0\to K_S\pi^0)=(34^{+27}_{-29})\%\;\;({\rm new})
\;. \label{inp}
\end{eqnarray}
The weak phase $\phi_1$ is set to $23^o$ from the time-dependent
$B\to J/\psi K^{(*)}$ measurement. Plus the unitarity relation
among the weak phases, all the above (primed and unprimed)
amplitudes, $\phi_2$ and $\phi_3$ can be solved exactly.

Neglecting the $O(\lambda^2)$ terms in the above parametrizations,
more than one $O(\lambda)$ solutions have been obtained excluding
the data labelled by ``new" \cite{Charng}. Some $O(\lambda)$
solutions from the $B\to\pi\pi$ data favor an amplitude $C$, which
is large and constructive to $T$, but some do not. Including the
new data, we are allowed to improve the accuracy up to
$O(\lambda^2)$, at which the approximate SU(3) flavor symmetry
relations,
\begin{eqnarray}
\frac{P}{T}\approx \frac{P'}{T'}\epsilon
e^{i\phi_1}\;,\;\;\;\frac{C}{T}\approx
\frac{C'}{T'}\;,\;\;\;\frac{P_{ew}}{P}\approx
\frac{P'_{ew}}{P'}\;, \label{su3}
\end{eqnarray}
with the factor $\epsilon\equiv \lambda^2/(1-\lambda^2)=0.05$,
come to help discriminate different solutions. This discrimination
is impossible at $O(\lambda)$, since $P_{ew}$ ($C'$) does not
appear in the $O(\lambda)$ parameterization for the $\pi\pi$
($K\pi$) modes. We argue, as pointed out in \cite{NK04,KMM03},
that it is unreasonable to apply the exact SU(3) flavor symmetry
to relate the amplitudes in the $\pi\pi$, $K\pi$ modes. On one
hand, the symmetry must be broken by QCD dynamics. On the other
hand, the amplitudes in Eqs.~(\ref{Mbpi1}) and (\ref{Mbpp1}) have
absorbed some subleading contributions through their
redefinitions. To be explicit, we have, even under the exact SU(3)
flavor symmetry,
\begin{eqnarray}
T-T'=T^a\;,\;\;\;\;C-C'=-T^a\;,\;\;\;\;
P-P'=P_{ew}^a\;,\;\;\;\;P_{ew}-P'_{ew}=P_{ew}^c\;,
\end{eqnarray}
with the electroweak penguin annihilation amplitude $P_{ew}^a$.
According to the power counting rules in \cite{Charng}, $C/T$ and
$C'/T'$ could differ by $O(\lambda)\sim 20\%$, and $P_{ew}/P$ and
$P'_{ew}/P'$ could differ by $O(\lambda^2)\sim 5\%$. Adding the
SU(3) symmetry breaking effect about 20\%-30\%, we assume that
Eq.~(\ref{su3}) may suffer corrections of order 30\%-50\%.

Considering only the central values of the data as a
demonstration, there are four exact solutions for the $K\pi$
modes:
\begin{eqnarray}
& &\frac{T'}{P'}=0.30e^{-170^oi}\;,\;\;\;
\frac{P'_{ew}}{P'}=0.13e^{7^oi}\;,\;\;\;
\frac{C'}{T'}=0.89e^{-24^oi}\;,\;\;\;\phi_3=63^o\;,
\label{ks1}\\
& &\frac{T'}{P'}=0.38e^{-8^oi}\;,\;\;\;
\frac{P'_{ew}}{P'}=0.42e^{93^oi}\;,\;\;\;
\frac{C'}{T'}=0.69e^{155^oi}\;,\;\;\;\phi_3=116^o\;,
\label{ks2}\\
& &\frac{T'}{P'}=0.55e^{-5^oi}\;,\;\;\;
\frac{P'_{ew}}{P'}=0.46e^{-90^oi}\;,\;\;\;
\frac{C'}{T'}=0.47e^{-160^oi}\;,\;\;\;\phi_3=116^o\;,
\label{on2}\\
& &\frac{T'}{P'}=0.85e^{-176^oi}\;,\;\;\;
\frac{P'_{ew}}{P'}=0.10e^{-163^oi}\;,\;\;\;
\frac{C'}{T'}=1.18e^{-179^oi}\;,\;\;\;\phi_3=58^o\;. \label{on3}
%& &\frac{T'}{P'}=0.26e^{-168^oi}\;,\;\;\;
%\frac{P'_{ew}}{P'}=0.17e^{34^oi}\;,\;\;\;
%\frac{C'}{T'}=1.01e^{-18^oi}\;,\;\;\;\phi_3=61^o\;,
%\label{ks1}\\
%& &\frac{T'}{P'}=0.28e^{-11^oi}\;,\;\;\;
%\frac{P'_{ew}}{P'}=0.37e^{98^oi}\;,\;\;\;
%\frac{C'}{T'}=0.94e^{171^oi}\;,\;\;\;\phi_3=118^o\;,
%\label{ks2}\\
%& &\frac{T'}{P'}=0.62e^{-5^oi}\;,\;\;\;
%\frac{P'_{ew}}{P'}=0.49e^{-89^oi}\;,\;\;\;
%\frac{C'}{T'}=0.42e^{-149^oi}\;,\;\;\;\phi_3=117^o\;. \label{on2}
\end{eqnarray}
Other solutions with $T'/P'>1$ or with all the phases different
from those in Eqs.~(\ref{ks1})-(\ref{on3}) by $180^o$ have been
suppressed. We have varied the data slightly around their central
values, and found that the above solutions are stable. Equation
(\ref{on2}), showing a large $P'_{ew}/P'$, is close to that
obtained from the $O(\lambda)$ analysis \cite{Charng}, which is
valid for a smaller $C'$. The other three with larger $C'/T'$ are
new, and can not be derived at $O(\lambda)$. We then input the
above $\phi_3$ values into the $B\to\pi\pi$ case, and look for
solutions satisfying Eq.~(\ref{su3}). Substituting
$\phi_2=180^o-\phi_1-\phi_3\sim 40^o$ from Eqs.~(\ref{ks2}) and
(\ref{on2}) into Eq.~(\ref{Mbpi1}), we find no solution,
indicating that Eqs.~(\ref{ks2}) and (\ref{on2}) can not be the
consistent solutions for both the $\pi\pi$ and $K\pi$ modes. In
fact, $\phi_2$ must be greater than $60^o$ in order for a solution
to exist. That is, the current $B\to\pi\pi$ data have imposed a
constraint on the allowed range of $\phi_2$. The phase $\phi_2\sim
90^o$ corresponding to Eqs.~(\ref{ks1}) and (\ref{on3}) gives four
solutions,
\begin{eqnarray}
& &\frac{P}{T}=0.41e^{150^oi}\;,\;\;\;
\frac{C}{T}=0.81e^{-55^oi}\;,\;\;\;
\frac{P_{ew}}{P}=0.18e^{12^oi}\;,\label{ss1}\\
& &\frac{P}{T}=0.41e^{150^oi}\;,\;\;\; \frac{C}{T}=0.77e^{-50^oi}
\;,\;\;\;\;\frac{P_{ew}}{P}=0.03e^{145^oi}\;, \label{ss2}\\
& &\frac{P}{T}=0.41e^{150^oi}\;,\;\;\; \frac{C}{T}=0.42e^{60^oi}
\;,\;\;\;\;\frac{P_{ew}}{P}=2.41e^{47^oi}\;, \label{ss3}\\
& &\frac{P}{T}=0.41e^{150^oi}\;,\;\;\; \frac{C}{T}=0.34e^{58^oi}
\;,\;\;\;\;\frac{P_{ew}}{P}=2.56e^{44^oi}\;. \label{ss4}
%& &\frac{P}{T}=0.38e^{150^oi}\;,\;\;\;
%\frac{C}{T}=0.81e^{-58^oi}\;,\;\;\;
%\frac{P_{ew}}{P}=0.22e^{10^oi}\;,\label{ss1}\\
%& &\frac{P}{T}=0.38e^{150^oi}\;,\;\;\; \frac{C}{T}=0.72e^{-47^oi}
%\;,\;\;\;\;\frac{P_{ew}}{P}=0.25e^{-174^oi}\;, \label{ss2}\\
%& &\frac{P}{T}=0.38e^{150^oi}\;,\;\;\; \frac{C}{T}=0.48e^{61^oi}
%\;,\;\;\;\;\frac{P_{ew}}{P}=2.5e^{49^oi}\;. \label{ss3}
\end{eqnarray}

It is easy to observe that only Eq.~(\ref{ss1}) obeys the
approximate relations to Eq.~(\ref{ks1}) shown in Eq.~(\ref{su3}):
$C/T$ in Eq.~(\ref{ss1}) differs from $C'/T'$ in Eq.~(\ref{ks1})
by $2|C/T-C'/T'|/(|C/T|+|C'/T'|)\sim 50\%$, and $P_{ew}/P$ differs
from $P'_{ew}/P'$ by about 30\%. For Eq.~(\ref{ss2}), the
direction of $P_{ew}/P$ is almost opposite to that of $P'_{ew}/P'$
in Eq.~(\ref{ks1}). For Eqs.~(\ref{ss3}) and (\ref{ss4}), the
magnitude of $P_{ew}/P$ is too much larger than that of
$P'_{ew}/P'$, and $C/T$ also dramatically differs from $C'/T'$.
Equation (\ref{on3}) is not favored, since none of
Eqs.~(\ref{ss1})-(\ref{ss4}) is close to it. As emphasized before,
a consistency like the one between Eqs.~(\ref{ks1}) and
(\ref{ss1}) means that the $B\to \pi\pi$, $K\pi$ data are really
consistent with each other! Note that Eqs.~(\ref{ks1}) and
(\ref{ss1}) correspond to the central values of the data. If
considering the allowed range, the two solutions can be even
closer.
%For example,
%varying the $B\to \pi\pi$ data within $1\sigma$,
%\begin{eqnarray}
%& &{\rm Br}(B^\pm\to \pi^\pm\pi^0) =5.7\times 10^{-6}\;,
%\nonumber\\
%& &{\rm Br}(B_d^0\to \pi^\pm\pi^\mp) =4.2\times 10^{-6}\;,
%\nonumber\\
%& &{\rm Br}(B_d^0\to \pi^0\pi^0) =1.43\times 10^{-6}\;,
%\nonumber\\
%& &{\cal A}(B^\pm\to \pi^\pm\pi^0) =-8\%\;, \nonumber\\
%& &{\cal A}(B_d^0\to \pi^0\pi^0) \approx 0\% \;,
%\end{eqnarray}
%and maintaining the others, we get
%\begin{eqnarray}
%\frac{P}{T}=0.38e^{150^oi}\;,\;\;\;
%\frac{C}{T}=0.81e^{-43^oi}\;,\;\;\;
%\frac{P_{ew}}{P}=0.26e^{54^oi}\;.
%\end{eqnarray}
%It is easy to find that the above $C/T$ differs from $C'/T'$ in
%Eq.~(\ref{ks1}) by $|C/T-C'/T'|/|C'/T'|\sim 35\%$, and that
%$P_{ew}/P$ does too from $P'_{ew}/P'$.

We conclude from our $O(\lambda^2)$ analysis:

$\bullet$ The extracted ratio $T'/P'$ in Eq.~(\ref{ks1}) is in
agreement with the theoretical prediction from the perturbative
QCD (PQCD) approach, $(0.20\pm 0.04)\exp(-156^oi)$ \cite{KLS,KS}.
The extracted $P/T$ in Eq.~(\ref{ss1}) becomes smaller than
$(0.77^{+0.58}_{-0.34})\exp[(137^{+14}_{-21})^oi]$ from the old
data \cite{ALP,BPRS}, and closer to the PQCD prediction
$(0.23^{+0.07}_{-0.05})\exp[(143\pm 5)^oi]$ \cite{KS,LUY}. The
extracted $P/T$ and $T'/P'$ are consistent with those obtained in
\cite{BFRPR}.

$\bullet$ The recent $\pi\pi$, $K\pi$ data do not imply a large
electroweak penguin amplitude, because of
$|P^{(')}_{ew}/P^{(')}|\approx 0.2$, contrary to the conclusion in
the literature \cite{BFRPR,WZ,Y03,BN,GR03}. The extracted
$P^{(')}_{ew}/P^{(')}$, consistent with the standard-model
estimation $(0.14^{+0.06}_{-0.05})\exp[(3^{+23}_{-18})^oi]$ quoted
in \cite{BFRPR}, shows no signal of new physics.

$\bullet$ The recent data imply a large color-suppressed tree
amplitude with $|C^{(')}/T^{(')}|\sim O(1)$ and with a
constructive interference between $C^{(')}$ and $T^{(')}$. Our
$C^{(')}/T^{(')}$ is in agreement with
$1.22^{+0.25}_{-0.21}\exp[-(71^{+19}_{-25})^oi]$ derived in
\cite{BFRPR}, but differs from that in \cite{CGRS} (\cite{HM04}),
which favors a larger (vanishing) relative strong phase between
$C^{(')}$ and $T^{(')}$. Contrary to $P^{(')}_{ew}/P^{(')}$, the
extracted $C^{(')}/T^{(')}$ is 4 times bigger than from the PQCD
prediction. Note that $C^{(')}$ represents an effective amplitude
in the parametrization, which contains additional contributions
%(for example, from the $u$ quark loop)
compared to that calculated
in PQCD.

$\bullet$ The extracted weak phases $\phi_2\sim 90^o$ and
$\phi_3\sim 60^o$ are consistent with those form the global
unitarity triangle fit \cite{Bona}. When the data of the
mixing-induced CP asymmetry in the $B_d^0\to \pi^0\pi^0$ modes
becomes available, $\phi_2$ and $\phi_3$ can be determined
independently from the $B\to\pi\pi$, $K\pi$ decays, respectively,
and the unitarity condition of the weak phases can be checked. The
criteria in Eq.~(\ref{su3}) will still apply to discriminate
different solutions.

$\bullet$ The hierarchy in Eqs.~(\ref{po}) and (\ref{pow}) is not
well respected by the extracted amplitude ratios. Nevertheless,
these ratios arise from the central values of the data, and their
ranges are expected to be as wide as found in \cite{Charng}. More
precise data are necessary for examining the power counting rules.

It should be stressed that the $c$-convention with the CKM matrix
element product $V_t=V_{td}V_{tb}^*$ being eliminated has been
adopted for the parametrization of the $B\to\pi\pi$ amplitudes in
\cite{BFRPR,ALP,BPRS}. Therefore, their definitions of the ratios
$P/T$ and $C/T$ differ from ours:
\begin{eqnarray}
\left.\frac{P}{T}\right|_{c}=\frac{(V_c/|V_t|)(P/T)|_{t}}
{1+(|V_u|/|V_t|)(P/T)|_{t}}\;,\;\;\;\;
\left.\frac{C}{T}\right|_{c}=\frac{(C/T)|_{t}-(|V_u|/|V_t|)(P/T)|_{t}}
{1+(|V_u|/|V_t|)(P/T)|_{t}}\;,
\end{eqnarray}
where the ratios $P/T$ do not involve the weak phases, the
subscript $c$ ($t$) denotes the $c$($t$)-convention, and the CKM
matrix element product $V_u$ is given by $V_u=V_{ud}V_{ub}^*$.
Because of the large relative strong phase between $P$ and $T$ in
Eq.~(\ref{ss1}), the magnitudes of the ratios in the
$c$-convention are larger than those in the $t$-convention by
$20\%\sim 30\%$, which does not affect the comparisons made above.

There are two reasons for the different conclusions drawn in this
work and in \cite{BFRPR,WZ,Y03}. First, if employing the SU(3)
flavor symmetry as in the above references, $ie.$, substituting
$C/T\approx 0.81e^{-58^oi}$ in Eq.~(\ref{ss1}) for $C'/T'$ in
Eq.~(\ref{Mbpp1}), and solving for other amplitudes, we obtain
$P'_{ew}/P'=0.44e^{-79^oi}$, close to
$0.36^{+0.52}_{-0.25}\exp[-(82^{+29}_{-36})^o]$ in \cite{BFRPR}.
It implies that the new physics signal may be a consequence of the
exact SU(3) flavor symmetry, an assumption which is too strong as
explained before \cite{NK04}. Second, if adopting the old data
\cite{Charng} for those labelled by ``updated", and solving
Eq.~(\ref{Mbpp1}), we derive
\begin{eqnarray}
\frac{T'}{P'}=0.26e^{-169^oi}\;,\;\;\;
\frac{P'_{ew}}{P'}=0.41e^{-85^oi}\;,\;\;\;
\frac{C'}{T'}=1.04e^{-51^oi}\;,\;\;\;\phi_3=78^o\;, \label{ks11}
\end{eqnarray}
in which $P'_{ew}/P'$ is also close to that in \cite{BFRPR}. It is
then realized that the recent data have exhibited the tendency
toward a small electroweak penguin amplitude. The SU(3) flavor
symmetry was not fully relaxed in \cite{WZ}: the strong phase of
$C/T$ remains equal for both the $\pi\pi$, $K\pi$ modes. The data
adopted in \cite{WZ} were not completely updated either, such as
the $B^\pm\to K^0\pi^\pm$ branching ratio $(21.8\pm 1.4)\times
10^{-6}$. Hence, it is not a surprise to conclude a large
electroweak penguin amplitude.

Other observations from the updated data include: the $B_d^0\to
K^\pm\pi^\mp$ modes involve only the penguin amplitude $P'$ and
the tree amplitude $T'$. Hence, the large CP asymmetry observed in
these modes confirms the large relative strong phase between $T'$
and $P'$, as predicted by the PQCD approach \cite{KLS}. The
$B^\pm\to K^0\pi^\pm$ branching ratio has increased and become
almost twice of the $B_d^0\to K^0\pi^0$ one, indicating that the
magnitude of the electroweak penguin amplitude $P'_{ew}$ needs not
to be large as shown in Eq.~(\ref{ks1}). If the large relative
phase between $T'$ and $P'$ is established, and $P'_{ew}$ is
small, the tiny CP asymmetry observed in the $B^\pm\to K^\pm\pi^0$
modes then implies an essential $C'$ \cite{Ligeti04}, whose effect
is to orient $T'+C'$ along $P'$ as shown in Eq.~(\ref{ks1}). A
large $P'_{ew}$ found in \cite{BFRPR} is also a possible solution
to the small CP asymmetry in the $B^\pm\to K^\pm\pi^0$ modes: the
effect of $P'_{ew}$ is to rotate $P'$, such that it bisects the
angle between $(T'+C')\exp(i\phi_3)$ and $(T'+C')\exp(-i\phi_3)$.
This solution, corresponding to Eq.~(\ref{on2}), however, has been
ruled out as shown above. Similarly, the possible large CP
asymmetry observed in the $B_d^0\to \pi^\pm\pi^\mp$ modes also
hints a large relative strong phase between $T$ and $P$. At last,
we check the configuration between $P_{ew}$ and $-(T+C)$, and find
that their ratio, excluding the CKM matrix elements, is given by
%$0.007\exp(25^oi)$ from Eq.~(\ref{ks1}) and by
$0.024\exp(5^oi)$ from Eq.~(\ref{ss1}). This ratio, being only the
central value, is roughly consistent with $0.013\exp(0^oi)$ from
the isospin symmetry \cite{NGW,NR98,GPY}.

We have performed an $O(\lambda^2)$ analysis based on the
topological-amplitude parametrizations for the $B\to \pi\pi$,
$K\pi$ decays. Combining the recent $\pi\pi$, $K\pi$ data, such an
investigation is allowed. We do not rely on the SU(3) flavor
symmetry, but only require the extracted amplitude ratios from the
$\pi\pi$, $K\pi$ modes to satisfy the approximate relations in
Eq.~(\ref{su3}). We have found that an exact solution from the
$B\to\pi\pi$ data and an exact solution from the $B\to K\pi$ data,
obeying this weaker and more reasonable requirement, indeed exist.
These solutions show a large color-suppressed tree amplitude
constructive to the tree amplitude, and a small electroweak
penguin amplitude. The corresponding weak phases $\phi_2\sim 90^o$
and $\phi_3\sim 60^o$ should be convincing due to the consistency
between the $\pi\pi$, $K\pi$ data. Compared to the predictions
from the PQCD approach, the extracted $P/T$, $T'/P'$, and
$P^{(')}_{ew}/P^{(')}$ are all understandable. Only
$C^{(')}/T^{(')}$, larger than the PQCD predictions, demands more
study. This discrepancy may be attributed to the different
definitions of $C^{(')}$ in this work and in the PQCD approach.
%enhancement from the nonfactorizable contributions to the
%color-suppressed tree amplitude as in the
%$B^0\to D^{(*)0}\pi^0(\rho^0)$ decays \cite{KKL,KeumP}.
An explicit next-to-leading-order evaluation will answer whether
this large ratio can be achieved. Because $C'$ is important, the
color-suppressed electroweak penguin amplitude $|P_{ew}^{'C}|\sim
0.22|C'|\sim 0.1|P'|$ \cite{GPY} might cause some minor effect,
which will be studied elsewhere. It has been also demonstrated
that the recent data move toward a small electroweak penguin
amplitude. Therefore, we intend to claim that there is no strong
signal of new physics from the $B\to \pi\pi$, $K\pi$ decays. Our
method provides a promising determination of the weak phases and
of the topological amplitudes, whose ranges allowed by the data
will be worked out in a forthcoming paper.

\vskip 1.0cm We thank H.Y. Cheng, C.W. Chiang, X.G. He, Y.Y. Keum,
A. Kundu, Z. Ligeti, S. Mishima, D. Pirjol, T. Yoshikawa, and Y.
Zhou for useful discussions. This work was supported by the
National Science Council of R.O.C. under the Grant No.
NSC-93-2112-M-001-014 and by the Taipei Branch of the National
Center for Theoretical Sciences of R.O.C..

\end{document}